\documentclass[reprint,aps,pra,twocolumn,floats,showpacs]{revtex4-1}

\usepackage{bm}
\usepackage{graphicx}
\usepackage{times}
\usepackage{hhline}
\usepackage{array}
\usepackage{amsmath}

\newcolumntype{C}[1]{>{\centering\arraybackslash}m{#1}}
\usepackage{color}

\begin{document}

\title{Fine structure of an exciton coupled to a single Fe$^{2+}$ ion in a CdSe/ZnSe quantum dot}

\author{T. \surname{Smole\'nski}}\email{Tomasz.Smolenski@fuw.edu.pl}
\author{T. \surname{Kazimierczuk}}
\author{M. \surname{Goryca}}
\author{W. \surname{Pacuski}}
\author{P. \surname{Kossacki}}

\affiliation{Institute of Experimental Physics, Faculty of Physics, University of Warsaw, ul. Pasteura 5, 02-093 Warsaw, Poland}

\date{\today}

\begin{abstract}
We present a polarization-resolved photoluminescence study of the exchange interaction effects in a prototype system consisting of an individual Fe$^{2+}$ ion and a single neutral exciton confined in a CdSe/ZnSe quantum dot. Maximal possible number of eight fully linearly-polarized lines in the bright exciton emission spectrum is observed, evidencing complete degeneracy lifting in the investigated system. We discuss conditions required for such a scenario to take place: anisotropy of the electron-hole interaction and the zero-field splitting of the Fe$^{2+}$ ion spin states. Neglecting of either of these components is shown to restore partial degeneracy of the transitions, making the excitonic spectrum similar to those previously reported for all other systems of quantum dots with single magnetic dopants.
\end{abstract}

\maketitle

Single quantum dot (QD) containing an individual transition metal ion provides an excellent opportunity to study the spin properties of a localized magnetic impurity interacting with the semiconductor lattice and confined band carriers, but isolated from any other ions in the sample \cite{Besombes_PRL_2004}. The variety of magnetic dopants with partially filled $3d$-shell that can be incorporated into QDs of different III-V or II-VI materials offers a wide range of combinations, which can be chosen in order to achieve desired properties of a magnetic-ion-QD system, e.g., long spin relaxation time \cite{kobak_nature_2014,Smolenski_JAP_2015,Varghese_PRB_2014, Goryca_PRB_2015}. This idea has been brought to life by the discovery of the fact that a single ion embedded inside the QD does not introduce any non-radiative excitonic recombination channels even for wide energy gap QDs \cite{kobak_nature_2014,Smolenski_JAP_2015}. This finding has extended the optical studies of the QDs doped with individual ions from the two classical systems of CdTe/ZnTe \cite{Besombes_PRL_2004,Leger_PRL_2005,Leger_PRL_2006,Goryca_PRL_2009,LeGall_PRL_2011,Trojnar_PRL_2011,LeGall_PRL_2009,Besombes_PRB_2014,Goryca_PRL_2014,Lafuente_PRB_2015,Smolenski_PRB_2015_x2m} and InAs/GaAs \cite{Kudelski_2007_PRL,Krebs_PRB_2009,Baudin_PRL_2011} QDs with single Mn$^{2+}$ ions to novel systems such as: CdSe/ZnSe QDs with single Mn$^{2+}$ \cite{kobak_nature_2014,Smolenski_PRB_2015,Smolenski_JAP_2015} and Fe$^{2+}$ \cite{Smolenski_Nature_2016} or CdTe/ZnTe QDs with single Co$^{2+}$ \cite{kobak_nature_2014,Kobak_arxiv_2016} and Cr$^{2+}$ \cite{Lafuente_PRB_2016,Lafuente_APL_2016,Lafuente_PRB_2017}. For all of these systems, despite different electronic configurations of the magnetic ions, the ion ground energy level was found to be an orbital singlet exhibiting an effective spin $S$. The degeneracy of $2S+1$ ion spin states is lifted by the interaction with strained semiconductor lattice as well as the spin-orbit coupling. In the leading order, the resulting magnetic anisotropy can be described by the term of $DS_z^2$, where $z$ corresponds to the QD growth axis. The value of $D$ parameter is relatively small ($\sim \mu$eV) for isoelectronic Mn$^{2+}$ ion \cite{LeGall_PRL_2009,Besombes_PRB_2014,Goryca_PRL_2014,Lafuente_PRB_2015} and a few orders of magnitude larger for the ions with a non-zero orbital momentum \cite{kobak_nature_2014,Smolenski_Nature_2016,Lafuente_PRB_2016,Papierska_PRB_2016,Kobak_arxiv_2016}. As such, in case of the ions like Fe$^{2+}$ or Cr$^{2+}$ only the subspace of some lowest-energy $|S_z|$ ion states is available at low temperatures.

The optical access to the subspace of thermally populated ion spin states is granted by the $s,p$-$d$ exchange interaction between the ion and a neutral exciton (X) confined inside the QD \cite{Gaj_book_2010}. Such interaction splits the excitonic energy levels, which gives rise to a presence of multitude of lines in the X photoluminescence (PL) spectrum \cite{Besombes_PRL_2004,Kudelski_2007_PRL,kobak_nature_2014,Smolenski_Nature_2016,Lafuente_PRB_2016}. The general character of this spectrum is, however, qualitatively different for a QD containing the ion with half-integer and integer spin $S$. In the former case, the Kramers rule protects the twofold degeneracy of the energy levels for both initial and final states of the excitonic recombination. As a consequence, the resulting X optical transitions are at least doubly degenerate, which does not allow to address the half-integer-spin states of the ion based solely on the energies of the X PL lines. Instead, the unequivocal readout of the half-integer-spin of the magnetic dopant requires to monitor both the energies and polarizations of the emitted photons, as experimentally demonstrated for the Mn$^{2+}$ and Co$^{2+}$ ions in II-VI QD systems \cite{Besombes_PRL_2004,kobak_nature_2014}. A fundamentally different physical picture holds for the magnetic ion with an integer spin, which does not warrant Kramers degeneracy. In this case, all of the X optical transitions may have different energies. This implies that, in principle, the PL spectrum of an optically active bright exciton (X$_\mathrm{b}$) of two possible spin configurations may consist of $2N^2$ distinct emission lines, where $N$ is the number of thermally populated integer-spin states of the magnetic ion. However, reaching this theoretical limit requires both the ion and the exciton spin states to be substantially mixed, which would allow the optical transitions between each initial and each final state of the X recombination. Up to now, such conditions have not been met for any QD systems doped with single integer-spin ions, as the number of experimentally observed excitonic emission lines in all of the previous studies \cite{Besombes_PRL_2004,kobak_nature_2014,Smolenski_JAP_2015,Varghese_PRB_2014,Goryca_PRB_2015,Leger_PRL_2005,Leger_PRL_2006,Goryca_PRL_2009,LeGall_PRL_2011,Trojnar_PRL_2011,LeGall_PRL_2009,Besombes_PRB_2014,Goryca_PRL_2014,Lafuente_PRB_2015,Smolenski_PRB_2015_x2m,Kudelski_2007_PRL,Krebs_PRB_2009,Baudin_PRL_2011,Smolenski_PRB_2015,Smolenski_Nature_2016,Kobak_arxiv_2016,Lafuente_PRB_2016,Lafuente_APL_2016,Lafuente_PRB_2017,Fainblat_NL_2016} was always well below $2N^2$.

Here we provide a direct experimental proof showing that the bright exciton PL spectrum from a QD with an integer-spin magnetic ion may feature all possible $2N^2$ optical transitions of different energies. This finding is revealed by our polarization-resolved PL studies of CdSe/ZnSe QDs doped with single Fe$^{2+}$ ions ($S=2$). Owing to a large Fe$^{2+}$ magnetic anisotropy \cite{Smolenski_Nature_2016}, this system is characterized by a relatively small subspace of the $N=2$ lowest-energy ion spin states with $S_z=\pm2$ that are available at low temperatures. We present experimental data showing that, in general, the X PL spectrum in such a QD indeed consists of eight distinct emission lines, each of them being fully (linearly) polarized as expected for nondegenerate transitions. Such a character of the excitonic spectrum is demonstrated to be due to the interplay of three different interactions: the $s,p$-$d$ exchange coupling between the ion and the X, the anisotropic electron-hole exchange mixing the two spin states of the bright exciton, and the spin-orbit coupling mixing the two $S_z=\pm2$ spin states of the Fe$^{2+}$ ion. We show that a simple spin Hamiltonian model including these interactions fully reproduces both the energies, relative oscillator strengths and polarization properties of the X emission lines. Additionally, we demonstrate that this model extended with the exciton and ion Zeeman terms provides also a correct description of the excitonic PL spectrum evolution in the magnetic field applied in the Faraday geometry.

Our experiments were performed on a sample grown by molecular beam epitaxy (MBE). It consisted of a single layer of self-organized CdSe:Fe QDs embedded in a ZnSe barrier \cite{Smolenski_Nature_2016}. The concentration of iron doping was adjusted to maximize the probability of finding a QD containing exactly one Fe$^{2+}$ ion. For the optical experiments, the sample was immersed in the pumped liquid helium ($T=1.8$~K) inside a magneto-optical cryostat equipped with a superconducting magnet producing a magnetic field of up to 10~T in the Faraday geometry. The PL was excited using a continuous-wave Ar-ion laser at 488~nm. A reflective-type microscope objective attached directly to the sample surface was used to focus the laser to a sub-micrometer spot and to collect the PL signal from the QDs. The collected PL was then resolved using either 0.5~m or 0.75~m monochromator and recorded with a CCD camera. A set of polarization optics placed in the signal beam (including a linear polarizer, $\lambda/2$ and $\lambda/4$ wave plates) enabled us to perform the PL measurements with either linear or circular polarization resolution.

\begin{figure}
\includegraphics{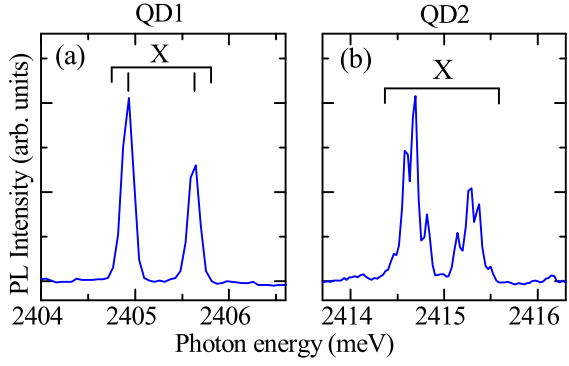}
\caption{PL spectra of the neutral exciton in two different CdSe/ZnSe QDs containing single Fe$^{2+}$ ions. The spectra were measured without polarization resolution. (a) The QD1 with negligible $\delta_\mathrm{Fe}$ splitting of the two Fe$^{2+}$ ground states. (b) The QD2 exhibiting large $\delta_\mathrm{Fe}$ splitting as well as strong anisotropic part of the electron-hole exchange interaction. Note that the spectral resolution of the QD1 spectrum is approximately 1.5 times lower than in the case of the QD2, which is due to shorter length of the used monochromator.}
\label{fig1_two_spectra}
\end{figure}

\begin{figure}
\includegraphics{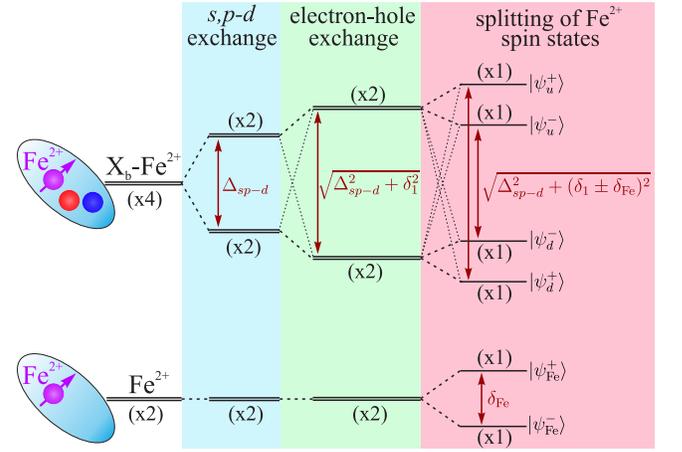}
\caption{The scheme illustrating the energies of both initial and final states of the bright exciton X$_\mathrm{b}$ recombination in a QD with a single Fe$^{2+}$ ion (note that level spacing is not in scale). Numbers in parentheses denote the degeneracy of the energy levels. For the sake of clarity, we omitted the higher-energy states of the Fe$^{2+}$ ion with $|S_z|<2$ that are not populated at helium temperatures.}
\label{fig2_energy_scheme}
\end{figure}

Our studies of the fine structure of the excitonic emission in a QD with a single Fe$^{2+}$ ion were performed on a number of randomly selected dots. The representative examples of the X PL spectra obtained for the two different Fe-doped QDs are presented in Figs~\ref{fig1_two_spectra}(a),(b). In the case of the QD1 [Fig.~\ref{fig1_two_spectra}(a)], the spectrum features a familiar two-line fine structure, which was previously established in Ref. \cite{Smolenski_Nature_2016}. The splitting between these lines arises mainly from the $s,p$-$d$ exchange interaction between the exciton and the resident ion. Due to a large magnetic anisotropy of the Fe$^{2+}$ ion as well as the heavy-hole character of the excitonic ground state in the QD, this exchange interaction is effectively Ising-like and takes a general form of $J_z\cdot S_z$, where $J_z$ and $S_z$ are the $z$-components of the excitonic and Fe$^{2+}$ spin operators (with $z$ being the quantization axis of both the hole and the Fe$^{2+}$ ion spins, i.e., the QD growth axis). As a result, the $s,p$-$d$ exchange splits four spin configurations $|J_z,S_z\rangle$ of the bright-exciton-Fe$^{2+}$ system into two subspaces consisting of states corresponding to antiparallel and parallel spin orientation of the X$_\mathrm{b}$ and Fe$^{2+}$ ion, i.e., $\left|\mp1,\pm2\right\rangle$ and $\left|\pm1,\pm2\right\rangle$, respectively. These two groups of states are separated by the energy of $\Delta_{sp-d}$ (see Fig.~\ref{fig2_energy_scheme}), which results in the presence of two distinct lines in the X PL spectrum. Such a character of the X emission remains unaltered also after taking into account the anisotropic part of the electron-hole exchange interaction $\delta_1$ \cite{Gammon_PRL_1996,Besombes_PRL_2000,Bayer_PRB_2002}, which is relatively strong in case of the studied CdSe/ZnSe QDs \cite{Puls_PRB_1999,Kulakovskii_PRL_1999,Patton_PRB_2003,Kobak_JPCM_2016}. The electron-hole exchange couples the two spin states $J_z=\pm1$ of the bright exciton associated with the same Fe$^{2+}$ ion spin projection $S_z$. Consequently, it increases the splitting between the two subspaces of the X$_\mathrm{b}$-Fe$^{2+}$ states to $\sqrt{\Delta_{sp-d}^2+\delta_1^2}$ (as depicted in Fig.~\ref{fig2_energy_scheme}), but does not lift the twofold degeneracy of these subspaces, thus preserving the two-line pattern of the excitonic emission.

\begin{figure*}
\includegraphics{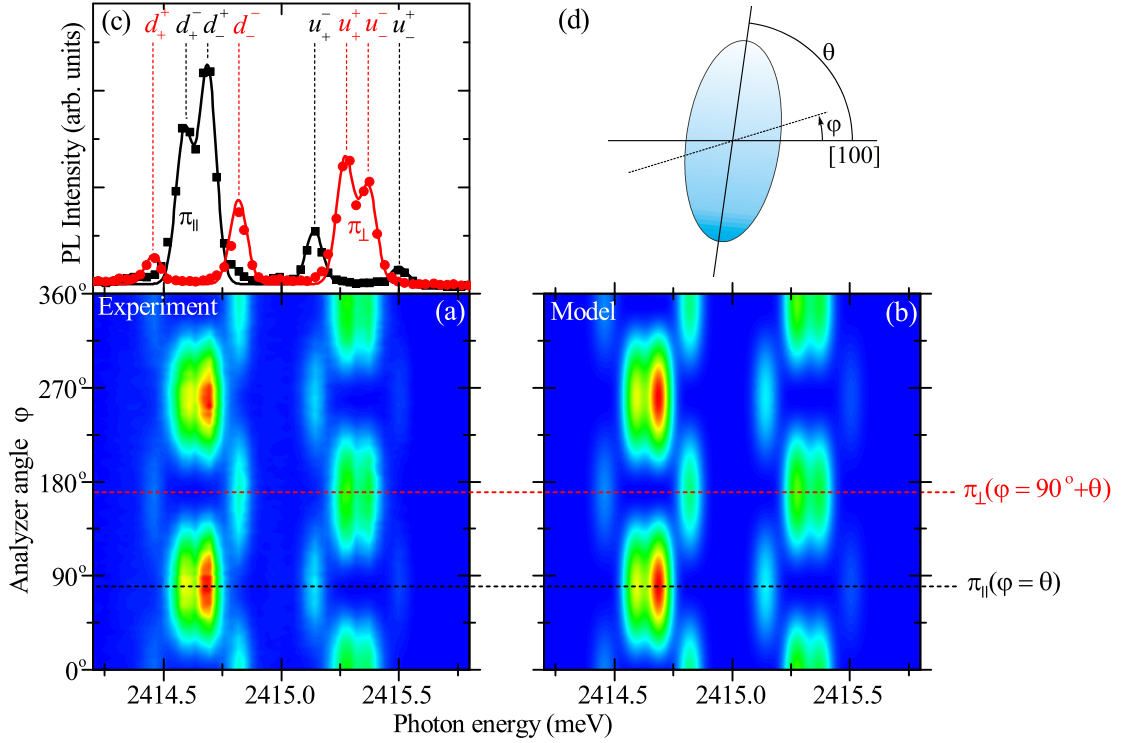}
\caption{(a) Color-scale map presenting the X PL spectra measured for the QD2 using different orientations of detected linear polarization. The polarization angle $\varphi$ is measured with respect to [100] crystallographic direction. For all spectra the continuous background was subtracted. Additionally, the subsequent spectra were corrected for a small intensity drift of the whole QD emission as well as residual spectral diffusion ($<0.03$~meV) using the reference unpolarized PL signal originating from the trion recombination in the same QD. (b) Theoretical simulation of the experimental data from (a) within the frame of a spin Hamiltonian model described in the text. The exchange energies used for the fit yield $\Delta_{sp-d}=0.53$~meV, $\delta_1=0.39$~meV, and $\delta_\mathrm{Fe}=0.23$~meV. (c) The X PL spectra detected in two orthogonal linear polarizations $\pi_\parallel$ (black) and $\pi_\perp$ (red or gray) oriented along the principal axes $\varphi=\theta$ and $\varphi=\theta+90^\circ$ of the electron-hole exchange interaction anisotropy (solid lines represent the model calculations). (d) Cartoon with definitions of the angles used in the discussion.}
\label{fig3_anizo_maps}
\end{figure*}

The physical picture turns out to be qualitatively different if, apart from the $s,p$-$d$ and electron-hole exchange interactions, there exists also the third interaction, which mixes the two Fe$^{2+}$ ion spin states $S_z=\pm2$. Such a mixing is characteristic of the magnetic ion with an integer spin, and for the Fe$^{2+}$ it arises due the higher-order spin-orbit coupling and a local in-plane anisotropy of the QD \cite{Smolenski_Nature_2016}. Regardless of its microscopic origin, such a mixing affects both initial and final configurations of the excitonic recombination. First, it splits the two Fe$^{2+}$ spin states in the final configuration by $\delta_\mathrm{Fe}$. Second, the invoked mixing of the ion states induces also a substantial reorganization of the X$_\mathrm{b}$-Fe$^{2+}$ energy levels leading, in general, to the presence of four nondegenerate states, as shown in Fig. \ref{fig2_energy_scheme}. Thus, for appropriately large value of $\delta_\mathrm{Fe}$ the X PL spectrum no longer consists of two emission lines, but exhibits more complex structure, as seen in the spectrum obtained for the QD2 [Fig.~\ref{fig1_two_spectra}(b)]. Interestingly, this spectrum still comprises of two broader spectral features separated mainly by the $s,p$-$d$ exchange interaction, but each of these features displays also a non-trivial internal fine structure.

An insight into the general properties of the bright X emission in a dot with a single Fe$^{2+}$ is provided by our linear-polarization-resolved PL measurements of the QD2. Figure~\ref{fig3_anizo_maps}(a) shows a series of the X PL spectra measured for different orientations $\varphi$ of detected linear polarization (where $\varphi=0^\circ$ corresponds to [100] crystallographic direction). The data clearly reveals the presence of eight distinct emission lines, which are fully linearly polarized along two orthogonal directions [the spectra detected in these two polarizations are presented in Fig.~\ref{fig3_anizo_maps}(c)]. This observation directly confirms that all four initial states of the X$_\mathrm{b}$-Fe$^{2+}$ system as well as the two final states of the Fe$^{2+}$ ion are indeed nondegenerate in the studied QD. Moreover, the optical transitions are possible between any given pair of these initial and final states. The resulting emission pattern consists of two quartets of lines separated by a pronounced central gap. For each of these quartets, the pair of stronger inner lines is orthogonally polarized to the pair of weaker outer lines. Furthermore, both the inner and outer pairs of lines within the lower-energy quartet are cross-polarized to the corresponding pairs of lines from the higher-energy quartet.

\begin{table*}
\begin{center}
\begin{tabular}{C{0.5cm}|C{4.5cm}|C{4.5cm}|}
\hhline{~--}
& $|\psi_\mathrm{Fe}^-\rangle$ & $|\psi_\mathrm{Fe}^+\rangle$ \\\hhline{-==}
\multicolumn{1}{|c||}{$|\psi_u^+\rangle$} & $\frac{1}{2}\left(1-\sin2\alpha_+\right)\cos^2(\varphi-\theta)$ & $\frac{1}{2}\left(1+\sin2\alpha_+\right)\sin^2(\varphi-\theta)$ \\\hline
\multicolumn{1}{|c||}{$|\psi_u^-\rangle$} & $\frac{1}{2}\left(1+\sin2\alpha_-\right)\sin^2(\varphi-\theta)$ & $\frac{1}{2}\left(1-\sin2\alpha_-\right)\cos^2(\varphi-\theta)$ \\\hline
\multicolumn{1}{|c||}{$|\psi_d^-\rangle$} & $\frac{1}{2}\left(1-\sin2\alpha_-\right)\sin^2(\varphi-\theta)$ & $\frac{1}{2}\left(1+\sin2\alpha_-\right)\cos^2(\varphi-\theta)$ \\\hline
\multicolumn{1}{|c||}{$|\psi_d^+\rangle$} & $\frac{1}{2}\left(1+\sin2\alpha_+\right)\cos^2(\varphi-\theta)$ & $\frac{1}{2}\left(1-\sin2\alpha_+\right)\sin^2(\varphi-\theta)$ \\\hline
\end{tabular}
\end{center}
\caption{Relative values of squared dipole moments of excitonic optical transitions in linear polarization oriented along a $\varphi$ direction for a QD with a single Fe$^{2+}$ ion [expressed using angles defined in Fig.~\ref{fig3_anizo_maps}(d)]. Each cell of the Table contains a value of $|\langle f|\mathcal{P}_\varphi|i\rangle|^2$ corresponding to a linearly-polarized transition between a given pair $|i\rangle$ and $|f\rangle$ of initial and final states of the excitonic recombination.\label{tab1}}
\end{table*}

All observed features of the excitonic emission can be quantitatively described within the frame of a simple spin Hamiltonian model, which includes three aforementioned interactions: the $s,p$-$d$ exchange, the electron-hole exchange, and the mixing of the two Fe$^{2+}$ ion spin states with $S_z=\pm2$. Using the following basis $({\left|+1,+2\right\rangle}$, ${\left|-1,+2\right\rangle}$, ${\left|+1,-2\right\rangle}$, ${\left|-1,-2\right\rangle})$ of the exciton-ion states $|J_z,S_z\rangle$, the Hamiltonian of the X$_\mathrm{b}$-Fe$^{2+}$ system can be expressed as
\begin{equation}
H=\frac{1}{2}
\begin{bmatrix}
\Delta_{sp-d} & \delta_1e^{-2i\theta} & \delta_\mathrm{Fe}e^{-2i\beta} & 0 \\
\delta_1e^{2i\theta} & -\Delta_{sp-d} & 0 & \delta_\mathrm{Fe}e^{-2i\beta}\\
\delta_\mathrm{Fe}e^{2i\beta} & 0 & -\Delta_{sp-d} & \delta_1e^{-2i\theta} \\
0 & \delta_\mathrm{Fe}e^{2i\beta} & \delta_1e^{2i\theta} & \Delta_{sp-d}
\end{bmatrix},
\label{eq:ham}
\end{equation}
where $\theta$ and $\beta$ denote the angles between [100] crystallographic direction and in-plane anisotropy axes of, respectively, the electron-hole exchange interaction and the mixing of the Fe$^{2+}$ ion states. Without loss of generality, in the following analysis we will assume that $\delta_1, \delta_\mathrm{Fe}\ge0$. Analytical diagonalization of the above Hamiltonian allows us to compute four eigenstates of the X$_\mathrm{b}$-Fe$^{2+}$ system. They are given by
\begin{eqnarray}
|\psi_u^\pm\rangle&=&\cos\alpha_\pm\left|P^\pm\right\rangle+\sin\alpha_\pm\left|A^\pm\right\rangle,\\
|\psi_d^\pm\rangle&=&\cos\alpha_\pm\left|A^\pm\right\rangle-\sin\alpha_\pm\left|P^\pm\right\rangle,
\label{eq:eigenstates_X}
\end{eqnarray}
where $\alpha_\pm$ stands for the mixing angles defined by
\begin{equation}
\tan(2\alpha_\pm)=\frac{\delta_1\pm\delta_\mathrm{Fe}}{\Delta_{sp-d}},
\label{eq:mixing_angles}
\end{equation}
while
\begin{eqnarray}
\left|P^\pm\right\rangle&=&\frac{1}{\sqrt{2}}\left[\pm e^{-i(\beta+\theta)}\left|+1,+2\right\rangle+e^{i(\beta+\theta)}\left|-1,-2\right\rangle\right],\quad \\
\left|A^\pm\right\rangle&=&\frac{1}{\sqrt{2}}\left[\pm e^{-i(\beta-\theta)}\left|-1,+2\right\rangle+e^{i(\beta-\theta)}\left|+1,-2\right\rangle\right],\quad
\end{eqnarray}
are the phase-rotated superpositions of the two $|J_z,S_z\rangle$ states corresponding either to parallel or antiparallel orientation of the exciton and the ion spins. The eigenenergies of the four eigenstates of the Hamiltonian $H$ read
\begin{eqnarray}
E\left(|\psi_u^\pm\rangle\right)&=&\frac{1}{2}\sqrt{\Delta_{sp-d}^2+(\delta_1\pm\delta_\mathrm{Fe})^2},\\
E\left(|\psi_d^\pm\rangle\right)&=&-\frac{1}{2}\sqrt{\Delta_{sp-d}^2+(\delta_1\mp\delta_\mathrm{Fe})^2}.
\label{eq:eigenenergies_X}
\end{eqnarray}
The above expressions provide a valuable insight into the properties of the excitonic states in Fe-doped QD. First, regardless of the values of the mixing terms $\delta_1$ and $\delta_\mathrm{Fe}$, the dominant admixtures to the higher-energy (lower-energy) eigenstates $|\psi_u^\pm\rangle$ $(|\psi_d^\pm\rangle)$ correspond to the basis states with parallel (antiparallel) spin alignment of the ion and the exciton. Second, since the mixing angles $\alpha_\pm$ depend solely on the strengths of the three considered interactions, the joint contributions of parallel (antiparallel) spin states to each of the eigenstates is independent of the in-plane anisotropy directions $\theta$ and $\beta$. In fact, the amplitude of these contributions scales with a sum or a difference of the mixing terms $\delta_1$ and $\delta_\mathrm{Fe}$ for the pairs of the eigenstates with outer energies ($|\psi_u^+\rangle$, $|\psi_d^+\rangle$) or inner energies ($|\psi_u^-\rangle$, $|\psi_d^-\rangle$), respectively. This statement is also valid in regard to the values of energy splittings between the invoked pairs of the eigenstates, which are increasing functions of $|\delta_1\pm\delta_\mathrm{Fe}|$, as seen from Eq.~(\ref{eq:eigenenergies_X}) and marked in Fig.~\ref{fig2_energy_scheme}.

To simulate the excitonic PL spectrum, we calculate the intensities and polarizations of the optical transitions between the above-introduced eigenstates of the X$_\mathrm{b}$-Fe$^{2+}$ system and the two $\delta_\mathrm{Fe}$-split eigenstates of the Fe$^{2+}$ ion in an empty dot, which are given by
\begin{equation}
|\psi_\mathrm{Fe}^\pm\rangle=\frac{1}{\sqrt{2}}\left[\pm e^{-i\beta}\left|+2\right\rangle+e^{i\beta}\left|-2\right\rangle\right],
\end{equation}
with the state corresponding to ``$+$'' being higher-energy. The PL intensity emitted in the transition $|i\rangle\rightarrow|f\rangle$ with a linear polarization oriented along a $\varphi$ direction is computed as a product of the initial state occupancy and the squared transition dipole moment $|\langle f|\mathcal{P}_\varphi|i\rangle|^2$, where $\mathcal{P}_\varphi$ is an interband linear polarization operator. This operator can be expressed as $\mathcal{P}_\varphi=(e^{i\varphi}\mathcal{P}_+-e^{-i\varphi}\mathcal{P}_-)/\sqrt{2}$ \cite{Koudinov_PRB_2004, tkaz_prb_2011}, where $\mathcal{P}_\pm$ is an interband operator for $\sigma^\pm$ polarized transition that annihilate the exciton occupying $|J_z=\pm1\rangle$ state. Using these definitions as well as the above expressions for the eigenstates, we obtain the values of $|\langle f|\mathcal{P}_\varphi|i\rangle|^2$  for the transitions between all pairs of initial and final states, which are listed in Table~\ref{tab1}. Each of these values take a general form of either $f_\mathrm{osc}\cos^2(\varphi-\theta)$ or $f_\mathrm{osc}\sin^2(\varphi-\theta)$, where $f_\mathrm{osc}$ denotes the corresponding (relative) oscillator strength. This result evidences that the excitonic transitions are indeed fully linearly-polarized along the two orthogonal directions $\theta$ and $\theta+90^\circ$. These directions are coincident with the principal axes of the electron-hole exchange interaction anisotropy, which demonstrates that the orientation $\beta$ of the Fe$^{2+}$ ion in-plane anisotropy has no influence on the optical spectrum.

Let us now focus on the oscillator strengths $f_\mathrm{osc}$ of the optical transitions. According to the Table~\ref{tab1}, each of them has a similar form $f_\mathrm{osc}=(1\pm\sin2\alpha_m)/2$, where $\alpha_m$ stands for one out of the two mixing angles $\alpha_\pm$. Given that $\alpha_\pm>0$ in the typical case of the experimentally studied QD2, we can identify the more (less) intense excitonic transitions with those exhibiting a ``$+$'' (``$-$'') sign in the expression for the oscillator strength. As such, there are two pairs of stronger transitions $|\psi_u^\pm\rangle\rightarrow|\psi_\mathrm{Fe}^\pm\rangle$ and $|\psi_d^\mp\rangle\rightarrow|\psi_\mathrm{Fe}^\pm\rangle$ labelled as $u^\pm_\pm$ and $d^\mp_\pm$, respectively. In agreement with the experimental data, both of these pairs are cross-linearly-polarized [see Fig.~\ref{fig3_anizo_maps}(c)]. The same polarization behavior holds also for the two pairs of weaker transitions $|\psi_u^\mp\rangle\rightarrow|\psi_\mathrm{Fe}^\pm\rangle$ and $|\psi_d^\pm\rangle\rightarrow|\psi_\mathrm{Fe}^\pm\rangle$, which are labelled in Fig.~\ref{fig3_anizo_maps}(c) as $u^\mp_\pm$ and $d^\pm_\pm$, respectively.

Notably, the fact that the excitonic PL lines originate from four different initial states hinders a direct comparison of their intensities. For example, the transitions $u^+_+$ and $d^+_-$ exhibit equal oscillator strengths, however the PL intensity emitted in the latter transition is clearly larger in the experiment, as seen in Fig.~\ref{fig3_anizo_maps}(c). The major factor contributing to this effect is non-equal occupancy of various initial states, which stems from a partial relaxation of the X$_\mathrm{b}$-Fe$^{2+}$ system during the excitonic lifetime. In order to account for this factor, we fit the experimental data in Fig.~\ref{fig3_anizo_maps}(a) with a following formula describing the excitonic PL intensity $I$ at a given emission energy $E$ and for the orientation $\varphi$ of detected linear polarization
\begin{multline}
I(E,\varphi)=\\\sum_{|i\rangle,|f\rangle}\rho\left(|i\rangle\right)|\langle f|\mathcal{P}_\varphi|i\rangle|^2g\left(E-E_0-E_{|i\rangle\rightarrow|f\rangle}\right),
\label{eq:sim_spectrum}
\end{multline}
where $\rho\left(|i\rangle\right)$ is a (fitted) occupancy of $|i\rangle$ initial state, $E_{|i\rangle\rightarrow|f\rangle}$ denotes the energy difference between $|i\rangle$ and $|f\rangle$ states, $E_0$ is an overall energy shift (bare exciton energy), while $g$ is a Gaussian profile of the width corresponding to the resolution of our experimental setup. Such a formula perfectly reproduces our experimental results including both the energies, polarization behavior, and relative intensities of all eight PL lines, as shown in Figs~\ref{fig3_anizo_maps}(b),(c). On this basis we determine all relevant parameters characterizing the studied QD2. As already inferred from our initial considerations, the $s,p$-$d$ exchange energy $\Delta_{sp-d}=0.53$~meV is found to dominate over the energies of the two mixing terms $\delta_1=0.39$~meV and $\delta_\mathrm{Fe}=0.23$~meV. Furthermore, coherently with the previous reports on II-VI QDs \cite{Leger_PRB_2007,tkaz_prb_2011,tkaz_prb_2013,Smolenski_PRB_2016,Kobak_JPCM_2016}, the linear polarization orientation $\theta$ defined by the electron-hole exchange interaction anisotropy is not coincident with any particular crystallographic directions. Finally and interestingly, the occupancies $\rho\left(|i\rangle\right)$ of different initial states are found to follow closely a Boltzmann distribution with an effective temperature of about 15~K. We stress that similar overall agreement between the experiment and the theoretical model was achieved for all studied QDs containing single Fe$^{2+}$ ions, which confirms that our conclusions concerning the general properties of the excitonic spectrum might be regarded representative for the studied QD system.

\begin{figure}
\includegraphics{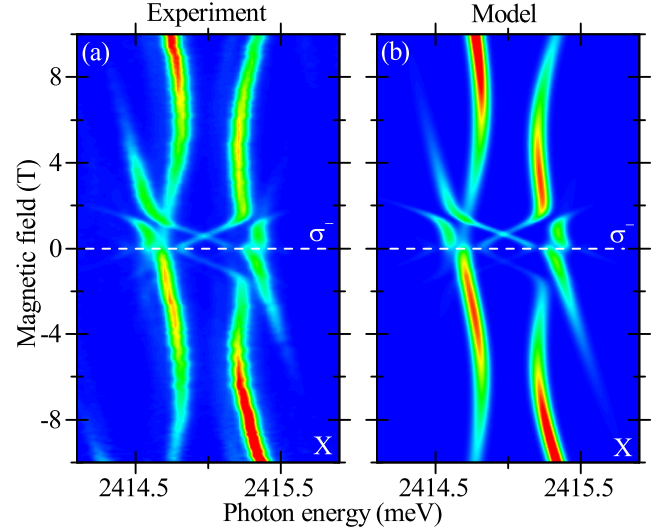}
\caption{(a) Color-scale map presenting the magnetic field dependence of the neutral exciton PL spectrum in the QD2 (the field was applied in the Faraday geometry, i.e., along the growth axis of the sample). The data were measured in $\sigma^-$ polarization of detection. (b) Corresponding simulation of the X magneto-PL performed within the frame of the spin Hamiltonian $H$ from Eq.~(\ref{eq:ham}) extended with a phenomenological diamagnetic shift term with $\gamma=0.65\ \mu$eV/T$^2$ as well as the ion and the exciton Zeeman terms with the corresponding g-factors equal to $g_\mathrm{Fe}=2.0$ and $g_\mathrm{X}=1.6$, respectively. The values of all the other model parameters were the same as in the zero-field fit from Fig.~\ref{fig3_anizo_maps}(b).}
\label{fig4_field}
\end{figure}

Having established the quantitative description of the excitonic PL spectrum in Fe-doped QD, it is interesting to examine how the properties of this spectrum evolve upon application of an external magnetic field in the Faraday geometry. The field introduces a Zeeman splitting between different spin states of both the exciton and the ion, which gives rise to a characteristic excitonic magneto-PL spectrum. Such a spectrum measured for the QD2 in $\sigma^-$ polarization of detection is shown in Fig.~\ref{fig4_field}(a). The general pattern is very similar to the one previously established in Ref.~\cite{Smolenski_Nature_2016}. This includes the presence of a distinctive cross-like feature, which is clearly visible in the range of small positive magnetic fields from $B=0$ to about 1.5~T. The two crossing lines correspond to the excitonic transitions involving the spin flip of the Fe$^{2+}$ ion from $S_z=\pm2$ to $S_z=\mp2$, which is partially allowed due to the mixing $\delta_\mathrm{Fe}$ of the ion spin states. The efficiency of this mixing is naturally enhanced when the effective magnetic field acting on the ion spin vanishes. This is realized either for $B=0$ in the absence of the exciton, or for a certain positive magnetic field (1.5~T in the case of the QD2), which compensates the $s,p$-$d$ exchange field coming from $|J_z=-1\rangle$ exciton emitting $\sigma^-$ polarized light. On this basis one would expect the crossing lines to appear only at positive magnetic fields, as indeed observed in Ref.~\cite{Smolenski_Nature_2016} for a relatively symmetric Fe-doped QD. However, in the case of the studied QD2, at least one out of the two crossing lines is also visible at negative fields. This is due to strong $\delta_1$ mixing of the two X spin states $|J_z=\pm1\rangle$ that at $B>0$ ($B<0$) is particularly efficient for the X states associated with $S_z=-2$ ($S_z=+2$) spin projection of the Fe$^{2+}$ ion, for which the excitonic Zeeman effect counteracts the ion-related exchange splitting. Importantly, all of the discussed details of the X magneto-PL can be described within the frame of our previously-developed theoretical model based on the spin Hamiltonian $H$ extended with an effective diamagnetic shift $\gamma B^2$ term as well as two Zeeman terms $g_\mathrm{Fe}\mu_BBS_z$ and $g_\mathrm{X}\mu_BBJ_z$ of the ion and the exciton, respectively. Using the same exchange energies determined previously from the zero-field analysis of the QD2, we are able to perfectly reproduce all the features observed in the experiment, as seen in the simulation of the X magneto-PL in Fig.~\ref{fig4_field}(b). The optical transitions in the presented simulation are calculated in the same manner as for the zero-field case described by Eq.~(\ref{eq:sim_spectrum}), except that now we use the $\mathcal{P}_-$ circular-polarization operator instead of a linear-polarization one, and include the thermalization of the Zeeman-split Fe$^{2+}$ ion spin states in the empty dot.

Concluding, we have thoroughly characterized the fine structure of the bright exciton PL spectrum in a CdSe/ZnSe QD with an individual Fe$^{2+}$ ion. In particular, we have demonstrated both experimentally and theoretically that, in general, this spectrum consists of eight distinct linearly-polarized emission lines, which correspond to optical transitions between all four initial states of the X$_\mathrm{b}$-Fe$^{2+}$ system and two final states of the Fe$^{2+}$ ion that are populated at helium temperatures. Such a rich spectral pattern was evidenced to be due to $\delta_\mathrm{Fe}$ mixing of the Fe$^{2+}$ spin states and the anisotropic part $\delta_1$ of the electron-hole exchange, which couples different spin states of the X that are split by the $s,p$-$d$ exchange with the magnetic ion. Based on our theoretical analysis we conclude that a necessary condition for the presence of a maximal number of eight lines in the X PL spectrum is a coexistence of both $\delta_\mathrm{Fe}$ and $\delta_1$ mixing interactions. Otherwise the spectral pattern is found to be much simpler, consisting of either two lines for vanishing $\delta_\mathrm{Fe}$ or four lines, when $\delta_1=0$, but $\delta_\mathrm{Fe}\neq0$. While the former pattern is observed experimentally for some Fe-doped QDs, the latter does not seem to be realized in the case of the studied CdSe/ZnSe QD system due to relatively strong electron-hole exchange coupling. Despite such a difference in possible spectral signatures of various Fe-doped QDs, their spectra are always found to be very similar upon application of a magnetic field of a few T in the Faraday geometry, which restores pure spin symmetry of the Fe$^{2+}$ eigenstates. This allows to conveniently address the Fe$^{2+}$ ion spin by optical means even in highly asymmetric QDs, which indicates that the QD anisotropy does not reduce the potential of the Fe$^{2+}$ ion as a two-level system of presumably long spin coherence time \cite{Smolenski_Nature_2016}.

\begin{acknowledgments}
This work was supported by the Polish National Science Centre under decisions DEC-2011/02/A/ST3/00131, DEC-2015/18/E/ST3/00559, DEC-2013/09/B/ST3/02603, and by the Polish Ministry of Science and Higher Education under project IP2015 031674. One of us (T.S.) was supported by the Polish National Science Centre through PhD scholarship grant DEC-2016/20/T/ST3/00028.
\end{acknowledgments}

%\bibliographystyle{apsrev_my}
%\bibliography{Fe_anizo_bibliography}

\end{document}